\documentclass{ws-procs9x6}
\newcommand{\open}{{<\kern -0.3 em{\scriptscriptstyle )}}}

\begin{document}

\title{Addressing transversity with \\
interference fragmentation functions}

\author{M. RADICI}

\address{Dipartimento di Fisica Nucleare e Teorica and \\
Istituto Nazionale di Fisica Nucleare, \\
v. Bassi 6, I27100 Pavia, Italy \\ 
E-mail: radici@pv.infn.it}


\maketitle

\abstracts{The class of interference fragmentation functions, arising from
interference among different hadron production channels, is reconsidered. Their 
symmetry properties with respect to chiral transformations 
allow building spin asymmetries where the quark transversity distribution can be
factorized out at leading twist. For the case of two leading spinless hadrons inside
the same current jet, the pair system is expanded in relative partial waves. The
cross section is represented on the helicity basis of the target and the fragmenting
quark, as well as on the relative orbital angular momentum of the pair. From the 
decay matrix being positive semi-definite, new bounds on the interference 
fragmentation functions can be derived. The expansion in partial waves allows to 
naturally incorporate in a unified formalism specific cases already studied in the 
literature, such as the fragmentation functions arising from the interference of 
two mesons in relative $s$ and $p$ waves, as well as the fragmentation of a spin-1 
hadron.}

\section{Introduction}
\label{sec:intro}

The interest in two-hadron fragmentation functions is justified by the search for a 
leading-twist mechanism capable to single out the chiral-odd transversity
distribution in an alternative and technically simpler way than the Collins 
effect~\cite{collins}. In fact, for semi-inclusive process where two leading 
spinless hadrons are produced inside the same current jet, the analysis of the 
leading-twist quark-quark correlator reveals a rich structure~\cite{io1}. 
Four fragmentation functions appear, related to various polarization states of the 
fragmenting quark. After integrating over the transverse momentum of the 
fragmenting quark, one of them survives, which is naive time-reversal odd 
(being related to the interference of different channels leading to the same final 
hadronic state; hence the naming convention of interference fragmentation function, 
IFF) and chiral odd; a single-spin asymmetry (SSA) can then be built to extract 
transversity at leading twist~\cite{io3}. In the SSA the critical parameter is just 
the azimuthal angle between the two-hadron plane and the laboratory plane; the 
insensitivity to the quark transverse momentum preserves collinear factorization, 
thus avoiding the introduction of Sudakov form factors that take into account the 
resummation of the soft-gluon radiation and possibly dilute the SSA as in the case 
of the Collins effect~\cite{boer-suda}. From numerical simulations, it turns out 
that the SSA are measurable, at least for the semi-inclusive deep-inelastic 
scattering (SIDIS)~\cite{io3}.

In this work we reanalyze the general quark-quark correlator for the fragmentation 
in two spinless hadrons in the helicity basis of the fragmenting quark and in the 
basis of the relative orbital angular momentum of the hadron pair. The advantage is 
twofold. First, a certain number of useful bounds on IFF can be deduced from the 
fact that the helicity matrices are positive semi-definite. Second, a general 
unifying formalism is deduced that naturally incorporates the specific case 
describing the interference between relative $s$ and $p$ waves of the hadron 
pair~\cite{jaffe}, as well as the case of spin-1 hadron 
fragmentation~\cite{ale-rho} in the subsector of the two-hadron relative $p$ wave. 

\begin{figure}[h]
\centerline{\epsfxsize=5cm\epsfbox{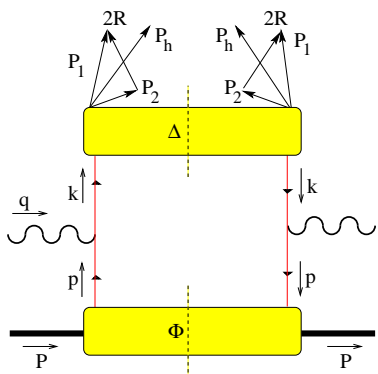}}   
\caption{The usual quark handbag diagram contributing at leading twist to the
semi-inclusive DIS in two leading hadrons, with total momentum $P_h=P_1+P_2$ and
relative momentum $R=(P_1-P_2)/2$. There is a similar diagram for anti-quarks.}
\label{fig:fig1}
\end{figure}

\vspace{-0.5truecm}

\section{The quark-quark correlator in the helicity basis}
\label{sec:q-q}

The semi-inclusive production of two spinless hadrons is depicted in 
Fig.~\ref{fig:fig1}. Inside a target nucleon with mass $M$, polarization $S$ and 
momentum $P$ ($P^2=M^2, S^2=-1, P\cdot S=0$), a virtual hard photon with momentum 
$q$ hits a quark with momentum $p$ promoting it to the state with momentum $k=p+q$. 
The quark then fragments into a residual jet and two leading unpolarized hadrons 
with masses $M_1, M_2,$ and momenta $P_1$ and $P_2$ ($P^2_1 = M_1^2, P^2_2 = 
M_2^2$), with $P_h=P_1+P_2$ ($P_h^2=M_h^2$) and $R=(P_1-P_2)/2$. We further define 
the light-cone fractions $x=p^+/P^+$, $z= P_h^-/k^-$, and $\zeta = 2R^-/P_h^-$, 
which describes how the total momentum of the hadron pair is split into the two 
single hadrons ($-1\leq \zeta \leq 1$ and $\zeta = 2\xi -1$, with $\xi$ defined in 
Ref.~\cite{io1}). On-shell conditions and the positivity requirement $R_T^2 \ge 0$
constrain the light-cone components of all the relevant 4-momenta.

Following the factorization hypothesis, the leading-twist cross section in the 
helicity basis (including polarizations of beam and target) can be written as
\begin{eqnarray}
   \frac{d^7\sigma}{d\zeta\;dM_h^2\;d\phi_R\;dz\;dx\;dy\;d\phi_S} &= &
   \sum\limits_{a} \; \rho_{\Lambda\Lambda'}(S) \, 
   [\Phi_a(x)]_{\chi_1\chi_1'}^{\Lambda\Lambda'} \, \left( 
   \frac{d\sigma^{eq_a}}{dy} \right)^{\chi_1\chi_1'\,;\,\chi_2\chi_2'} \nonumber \\
 & &\qquad [\Delta_a(z,\zeta,M_h^2,\phi_R)]_{\chi_2 \chi_2'} \; , 
\label{eq:crossh}
\end{eqnarray}
where $\phi_R, \phi_S$ are the azimuthal angles of the transverse components of $R$ 
($\vec R_T$) and of $S$ ($\vec S_T$) with respect to the lepton scattering plane, 
respectively. 

In Eq.~(\ref{eq:crossh}), $\rho_{\Lambda\Lambda'}(S)$ is the target helicity 
density matrix (with $\Lambda,\Lambda'$ the helicities of the nucleon legs 
in Fig.~\ref{fig:fig1}). The q-q correlator $\Phi_a(x)$ describes the lower blob 
in Fig.~\ref{fig:fig1} and contains the well known momentum $f_1^a$, helicity
$g_1^a$ and transversity $h_1^a$ distributions for the flavor $a$. The indices 
$(\chi_1,\chi_1')$ identify the parton helicities for the emerging quark legs in 
Fig.~\ref{fig:fig1}. The matrix satisfies general requirements, in particular it is 
positive semidefinite, from which the well known Soffer bound is obtained. 

Then, $\left( d\sigma^{eq_a}/dy\right)^{\chi_1\chi_1'\,;\,\chi_2\chi_2'}$ 
represents the standard elementary electron-quark cross section for a flavor $a$
with $y=E-E'/E$ the beam energy fraction delivered to the target. Finally, the q-q 
correlator $\Delta_a(z,\zeta,M_h^2,\phi_R)$ refers to the upper
blob in Fig.~\ref{fig:fig1} and is described as 
\begin{equation}
  [\Delta_a(z,\zeta,M_h^2,\phi_R)]_{\chi_2 \chi_2'} 
     = \frac{1}{2} \, \left( \begin{array}{cc}
       D_1^a & i e^{i\phi_R} \frac{|\vec R_T|}{M_h} 
       H_1^{\open\,a} \\[5pt]
       -i e^{-i\phi_R} \frac{|\vec R_T|}{M_h} H_1^{\open\,a} & 
       D_1^a  \end{array} \right) \; .
\label{eq:delta}
\end{equation}
The function $D_1^a(z,\zeta,M_h^2)$ represents at leading twist the probability that
two spinless hadrons are produced with $M_h$ invariant mass and carrying a $z$
fraction of the parent fragmenting quark with flavor $a$, sharing it in 
$(1+\zeta)/2$ and $(1-\zeta)/2$ parts inside the pair. The same interpretation can
be ascribed to $H_1^{\open\,a}$ but for transversely polarized quarks. The 
following bounds derive:
\begin{equation}
   D_1^a(z,\zeta,M_h^2) \geq 0 \qquad D_1^a(z,\zeta,M_h^2) \geq 
   \frac{|\vec R_T|}{M_h} |H_1^{\open\,a}(z,\zeta,M_h^2)| \; .
\label{eq:ffbounds}
\end{equation}
The function $H_1^{\open\,a}$ is the only naive $T$-odd IFF surviving the 
integration upon $\vec k_T$; it is also chiral odd and can be used to isolate the 
transversity $h_1^a$ by considering the SSA for unpolarized lepton beam and 
transversely polarized target~\cite{io3}.


\begin{figure}[h]
\centerline{\epsfxsize=2.8cm\epsfbox{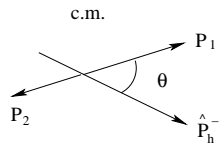}}   
\caption{The hadron pair in the cm frame. The light-cone direction $P_h^-$
identifies the current jet, while $\theta$ is the cm polar angle of the pair.}
\label{fig:fig2}
\end{figure}


\vspace{-0.5truecm}

\section{Partial-wave expansion}
\label{sec:lm}

If the invariant mass $M_h$ of the two hadrons is not very large, the pair can be
assumed to be produced mainly in the relative $s$- or $p$-wave channels in the cm
frame, which is defined by the back-to-back emission of the two hadrons. The 
direction identified by this emission forms an angle $\theta$ with the dominating
light-cone direction $P_h^-$ (see Fig.~\ref{fig:fig2}). The crucial remark is that 
in this frame $\zeta = a + b \cos \theta$, with $a,b$, functions only of the 
invariant mass. This suggests to expand the IFF in Eq.~(\ref{eq:delta}) onto the 
basis of Legendre polynomials in $\cos\theta$, namely 
\begin{eqnarray}
   D_1\big(z,\zeta(\cos\theta),M_h^2\big) &= &\sum\limits_{n} \, D_{n}(z,M_h^2) \, 
   P_n (\cos\theta) \nonumber\\
   H_1^{\open}\big(z,\zeta(\cos\theta),M_h^2\big) &= &\sum\limits_{n} \, 
   H_{n}^{\open}(z,M_h^2) \, P_n (\cos\theta)  \; .
\label{eq:fflm}
\end{eqnarray}
Here, the expansion includes the first three terms only $(n\leq 2)$, 
which are the minimal set required to describe all the ``polarization'' states of 
the system in the cm frame for relative partial waves $L=0,1$; they will be
conveniently indicated by the pair of indices $(i,j)$, with $i,j=O,L,T$ for
unpolarized $s$-wave, longitudinal and transversely polarized $p$-wave amplitudes. 
After inserting the expansion~(\ref{eq:fflm}) in the correlator~(\ref{eq:delta}), 
it is also useful to project out of the obtained matrix the information about the
orbital angular momentum of the system encoded in the angular distribution. In fact, 
for $L\leq1$ the decay matrix for the spinless hadron pair is given by specific
bilinear combinations of spherical harmonics as ${\mathcal D}_{MM'}^{LL'}
(\theta,\phi_R) = 4\pi\, Y_{LM} \, Y^*_{L'M'}$, from which the q-q
correlator~(\ref{eq:delta}) can be decomposed as
\begin{equation}
  [\Delta(z,\zeta,M_h^2,\phi_R)]_{\chi_2 \chi_2'} = 
    [\Delta(z,M_h^2)]_{MM'\,\chi_2\chi_2'}^{LL'} \;
    {\mathcal D}_{M'M}^{L'L}(\theta,\phi_R) \; ,
\label{eq:delta-open}
\end{equation}
where $[\Delta(z,M_h^2)]_{MM'\,\chi_2\chi_2'}^{LL'}$ contains the various $D_{ij}$ 
and $H^{\open}_{ij}$ components and, being positive semi-definite, gives their 
corresponding bounds.

Using Eq.~(\ref{eq:delta-open}) inside Eq.~(\ref{eq:crossh}), we can take advantage 
of the full power of the analysis in the helicity formalism. It is particularly 
interesting to consider the case for an unpolarized beam and a transversely 
polarized target, {\it i.e.} 
\begin{equation}
  \frac{d^7\sigma_{OT}}{d\zeta\;dM_h^2\;d\phi_R\;dz\;dx\;dy\;d\phi_S} \propto 
  \sin (\phi_R+\phi_S)\, 
   h_1^a(x)\, \sin\theta \, \Big[ H_{OT}^{\open}+\cos\theta H_{LT}^{\open} \Big] 
   \; ,
\label{eq:crosshlmOT}
\end{equation}
because we can see that the transversity $h_1$ can be matched by two different
chiral-odd, naive $T$-odd IFF: one pertaining the interference between $s$- and
$p$-wave states of the hadron pair, $H_{OT}^{\open}$ (corresponding to the 
hypothesis first formulated in Ref.~\cite{jaffe}), the other pertaining the
interference of the system in the $p$ wave only, $H_{LT}^{\open}$ (naturally linked 
to the analysis developed in the case of a spin-1 hadron 
fragmentation~\cite{ale-rho}). As for the former, it is enough 
to take $\phi_S=0$ in Eq.~(\ref{eq:crosshlmOT}), to integrate the $\theta$ 
dependence away in one emisphere and to realize that the azimuthal angle defined 
in Ref.~\cite{jaffe} is $\phi=\textstyle{1\over 2}\pi-\phi_R$. However, as it is 
evident from Eq.~(\ref{eq:crosshlmOT}) itself, the most general approach leads to an 
unfactorized $(z,M_h^2)$ dependence of the fragmentation function and to a 
completely different behaviour of the single-spin asymmetry with respect to 
Ref.~\cite{jaffe} (for a quantitative comparison, see Ref.~\cite{io3}). 

Finally, it is also possible to generalize the previous analysis to the case of
unintegrated transverse momenta. The formulae, however, are more involved and will
be shown elsewhere~\cite{ioale}.

\section*{Acknowledgments}

The author gratefully acknowledges the collaboration with A. Bacchetta, on which
this work is based.

\end{document}